# Tunneling Anisotropic Magnetoresistance in Co/AlO$_x$/Au Tunnel Junctions

R. S. Liu[a,b], L. Michalak[c], C. M. Canali[c], L. Samuelson[b] and H. Pettersson[a,b]*

[a] *Center for Applied Mathematics and Physics, Halmstad University, Box 823, SE-301 18 Halmstad, Sweden*

[b] *Solid State Physics/ the Nanometer Structure Consortium, Lund University, Box 118, SE-22100 Lund, Sweden*

[c] *Department of Physics and Mathematics, School of Pure and Applied Natural Sciences, Kalmar University, 391 82 Kalmar, Sweden.*

*Corresponding author. E-mail:  Hakan.Pettersson@ide.hh.se

**Abstract**

We observe spin-valve-like effects in nano-scaled thermally evaporated Co/AlO$_x$/Au tunnel junctions. The tunneling magnetoresistance is anisotropic and depends on the relative orientation of the magnetization direction of the Co electrode with respect to the current direction. We attribute this effect to a two-step magnetization reversal and an anisotropic density of states resulting from spin-orbit interaction. The results of this study points to future applications of novel spintronics devices involving only one ferromagnetic layer.

The conventional tunneling magnetoresistance (TMR), related to the relative orientation of the magnetization of two ferromagnetic electrodes, is defined as $TMR = \frac{R_{AP} - R_P}{R_P}$, where $R_A$ and $R_{AP}$ are the device resistances in the parallel and antiparallel magnetization configuration, respectively. TMR has been extensively investigated both experimentally and theoretically because of promising applications, e.g. magnetic random access memories (MRAM) and magnetic sensors[1]. A lot of research has focused on spin transport studies of ferromagnetic/nonmagnetic material/ferromagnetic (F/N/F) sandwich structures, where the two ferromagnetic electrodes act as the spin injector and spin detector respectively[2-4]. In order to draw correct conclusions on e.g. the spin relaxation time in such devices it is obviously crucial to clarify that the observed TMR is genuine, and that it results from changes in the relative magnetization between the injecting and detecting electrodes. In fact, a spin-valve-like effect can occur also in tunnel devices where only one electrode is ferromagnetic. Indeed, recent transport experiments on normal metal/single-wall carbon nanotube/ferromagnetic semiconductor tunnel devices[3] and normal metal/insulator/ferromagnetic semiconductor devices[5] display TMR-like signals. In the latter case the TMR depends strongly on the orientation of the magnetization with respect to the current direction or crystal axes, and the expression *tunneling anisotropic magnetoresistance* (TAMR) has been coined to describe this effect. Following these first observations, TAMR effects have also been reported in (Ga,Mn)As nanoconstrictions[6], Fe/GaAs/Au epitaxial metal semiconductor system[7] and predicted for other tunnel junction devices with a single magnetic electrode of various ferromagnetic materials[8,9]. It has been proposed that TAMR might be a general effect observable in other magnetic materials and devices[6]. Note that even before the discovery of TAMR, Bode *et al*[10] found from scanning tunneling microscope (STM) experiments that the electronic structure of thin Fe film on W(110) depends on the magnetization direction, implying that TMR-like signals can be observed by orienting the magnetization direction of a single ferromagnetic electrode.

In this paper, we report on transport studies of nano-scaled ferromagnetic Co/AlO$_x$/Au/AlO$_x$/Co single electron transistors (F-SETs). We find considerable TMR signals which depend on the direction of the applied external magnetic field. From careful reference measurements on single Co/AlO$_x$/Au tunnel junctions, we conclude that the anisotropic TMR observed in the F-SETs is in fact a TAMR effect originating from a single dominating Co/AlO$_x$/Au junction. We attribute this TAMR effect to the dependence of the spin-orbit-interaction-induced anisotropic density of states (DOS) on the magnetization orientation with respect to current direction in the Co electrode. Our results underline the importance of careful reference measurements to avoid misinterpretations of spin-dependent transport experiments.

We fabricate our devices by employing a high-precision alignment procedure invoked during the electron beam lithography[11]. The devices sit on top of a 300 nm thick SiO$_2$ layer, thermally grown on a Si substrate. Fig 1 (a) shows a Co/AlO$_x$/Au/AlO$_x$/Co F-SET with the two ferromagnetic electrodes coupled to the central Au island by AlO$_x$ tunnel barriers. The wire-like central Au island measures 150 nm in length, 20 nm in width and 25 nm in thickness. The islands are prepared together with Au side-gates in one step employing electron-beam lithography, followed by thermal evaporation of Au and an approximately 1.8 nm thick layer of Al. Tunnel barriers of AlO$_x$ are subsequently formed in an ozone ambience with an oxygen gas flow of 500 cm$^3$/min for 3.5 minutes. Ferromagnetic Co source and drain electrodes, 40 nm thick, are defined on top of the Au islands using a high-precision alignment procedure in a second electron beam lithography step. The area of the tunnel junctions amounts to only about 40 × 20 nm$^2$. The source electrode has a length of 1.5 μm and a width of 80 nm. The corresponding dimensions of the drain electrode are 800 nm and 280 nm, respectively. The separation between the parallel drain and source electrodes is approximately 55 nm. Because of shape anisotropy, the two electrodes are expected to undergo magnetic reversal at different magnetic fields[12]. After the fabrication, electrical conductance measurements are carried out at 4.2K in a liquid helium Dewar. Fig 1(b) shows a typical non-linear current-voltage (I-V) curve for the device at 4.2K. The modulation of the current with gate bias is shown in the inset at different drain bias, with each peak corresponding to addition of one electron to the island. Following these measurements, the sample is transferred to a cryostat housing a 6T superconducting magnet where magnetoresistance measurements are carried out at low temperature. The magnetic field is in the plane of the device with a tunable orientation with respect to the orientation of the electrodes.

Figure 2(a) and 2(b) show the typical dependence of the resistance of a $Co/AlO_x/Au/AlO_x/Co$ F-SET on the magnetic field for two sweep directions, with the in-plane field applied parallel and perpendicular to the long axis of the Co electrodes, respectively. In both figures the resistance traces exhibit spin-valve-like behavior with inverse MR (i.e. a dip in the resistance at small field strengths) and normal MR (peak in the resistance at low field strengths) occurring respectively for the two cases. Comparing Figs. 2(a) and (b) it is evident that the width of the MR traces is broader for perpendicular magnetic field. The inverse MR observed in Fig. 2(a) stands in sharp contrast to conventional TMR frequently reported for F-SETs since an anti-parallel orientation of the magnetization direction of the two leads is expected at low field strengths, and thus a peak in the resistance.

To clarify whether the observed MR signal indicates a spin-polarized current transport via the central Au island, we fabricated test samples with a T-shaped central Au island as shown in the inset of Fig. 3(a) using the same fabrication procedure as described above. This sample design enables us to independently measure the MR of the double tunnel junctions (i.e. the F-SET) and either individual tunnel junction. Figs. 3(a), (b) and (c) show the characteristic MR of the double tunnel junction (Co1/AlO$_x$/Au/AlO$_x$/Co2) and the single tunnel junctions (Au/AlO$_x$/Co1 and Au/AlO$_x$/Co2), respectively, with the magnetic field applied parallel to the long (easy) axis of the Co leads. The MR traces in Fig. 3(a) and 3(b) are very similar both with respect to shape, magnitude and sign of the signal. The MR trace in Fig. 3(c) is quite different. From the Figs 3 a-c it is evident that the MR signal observed for the F-SET in fact stems from the tunnel junction exhibiting the largest resistance. We point out here that this example clearly shows the necessity of carefully verifying the origin of a spin-valve signal in experiments involving injection and detection of spin polarized current.

In order to eliminate any dipole-dipole interactions between the magnetic moments of the two ferromagnetic electrodes[13], we fabricated Co/AlO$_x$/Au structures consisting of one ferromagnetic electrode (Co) coupled to a nonmagnetic metal (Au) electrode via an AlOx tunnel barrier, as shown in the left insets of Fig 4 (a) and (b). Fig 4 (a) and (b) show the resistance traces versus magnetic field applied parallel to and perpendicular to the long axis of the Co lead, respectively. The inverse and normal tunneling MR, reaching large magnitudes of 30% - 40%, can clearly be observed in the two figures. The dependence of the MR on the applied bias for both cases are shown in the right insets from which a gradual decrease of the signal is observed when increasing the bias from 10 mV to 75mV. We mention that we have investigated in total 7 Au/AlO$_x$/Au devices, none of which exhibited any MR behavior. We thus conclude that the MR signal in the Co/AlO$_x$/Au junctions is due to the single ferromagnetic Co electrode.

The transport properties of the junctions described above can be interpreted in terms of the recently discovered TAMR[5, 6]. The TAMR in tunnel junctions with one ferromagnetic lead reflects the dependence of the tunneling density of states on the direction of the magnetization of the magnetic lead with respect to the direction of the current[6]. As for the normal anisotropic magnetoresistance (AMR) [14, 15] in ferromagnetic metals, the origin of the anisotropy is to be found in the spin-orbit (SO) interaction.

In ferromagnetic metals, the spin-orbit (SO) interaction mixes the spin of the exchange-split d-bands responsible for the magnetic state, and their orbital character, lifting possible degeneracies at the Fermi level present when the SO is absent. The density of states of these d-bands at the Fermi levels depends on the direction of the magnetization, which can be manipulated with an external magnetic field. On the other hand, the transport properties of the metal depends on free-particle-like states belonging to weakly polarized s-bands. The resistivity of the metal reflects the scattering of electrons from conductive s states into (out from) localized d-states, which occurs via various types of impurity potentials. The scattering amplitude of these processes depends crucially on the wave vector of the s-states involved. It turns out[14] that when the wave-vector of the conductive s-states and therefore the current is parallel to the magnetization, the resulting resistivity is larger than for the case in which the current and the magnetization direction are orthogonal. This is the normal AMR in ferromagnetic metals[14,15]. We will see below that our experimental tunneling MR results bear the signatures of this effect, thus establishing the origin of the observed TAMR.

In Fig. 4(a), the magnetization of the Co electrode is parallel to the current direction at large positive or negative field strengths. Increasing or decreasing the magnetic field in this region has very little effect on the resistance, showing that the normal isotropic MR in our ferromagnetic sample is very small. The magnetization of the electrode will reverse as the magnetic field is swept from negative fields to positive fields (or vice-versa). For magnetic structures smaller than a few times the magnetic exchange length, $\lambda_{ex} = \sqrt{4\pi A / \mu_0 M_S^2} \approx 2$ nm for Co, ($M_S$ is the saturation magnetization, A is the exchange stiffness[16]), the magnetization reversal process is expected to be coherent, as described by the Stoner-Wohlfarth model[17]. Since the width of the Co electrode in our devices is much larger than the magnetic exchange length[18] the magnetization reversal process most likely takes place through domain walls propagating through the structure[19]. Specifically we mean that the switching of the magnetization direction as the field is swept towards small field strengths is mediated by the nucleation of domain walls, which initially re-orients the magnetization of the Co electrode by 90° to become perpendicular to the current direction[20]. In the switching region we first observe an abrupt decrease of the resistance. In fact, this is exactly what is expected on the basis of the theory of the normal AMR, which predicts lower resistance when the current is orthogonal to the magnetization. As the external field increases (in opposite direction), the magnetization direction rotates by an additional 90° through a corresponding domain wall annihilation to complete the reversal. The magnetization is now again parallel to the current and the resistance increases back and does not change again appreciably when the field is further increased outside the switching region. A complementary scenario explains the results of Fig. 4 (b). Here a perpendicular orientation of the magnetization with respect to the current direction at large field strengths gives rise to "low" resistance. In the switching region at small field strengths the magnetization has most likely an orientation parallel to the current and we expect an increase in the resistance, followed by a quick drop when the switching is completed and the magnetization returns perpendicular to the current. Our data are hence fully consistent with this picture. It is interesting to note that the step-like resistance changes always occurs at smaller field strengths for a setting with parallel orientation of the field with respect to the current. This can readily be understood since the long axis of the electrode is the easy axis, and the magnetic anisotropy energy favors a parallel configuration. It should be noted that the switching field varies from sample to sample, indicating that the magnetic domain texture varies from sample to sample. We have investigated 12 devices, yet only 4 of them display such clear TAMR as shown in Fig. 4, indicating that the quality of the Co/AlO$_x$ interface plays an important role.

Our experimental results on TAMR show a similarity to normal AMR[6], suggesting that their physical origins are related and both being due to the SO interaction. We therefore ascribe the anisotropic TMR of our devices to the dependence of the tunneling density of states at the Fermi energy on the magnetization orientation with respect to the current direction, similar to what was recently found in other hybrid tunnel systems [5, 6]. Looking at Fig.4, the tunnelling current should be orthogonal to the plane of the flat Co electrode and, as such, should be fairly unaffected by the direction of the magnetization when the latter rotates in that plane. While we cannot complete exclude this possibility, our analysis and understanding of the device fabrication suggests an alternative interpretation. In the fabrication procedure the central Au island is constructed first followed by forming of the $AlO_x$. The subsequently deposited Co electrode is not suspended *on top* of the $Au/AlO_x$ island but is rather *wrapped* around it i.e. the island actually penetrates into the Co electrode. This implies that the current can tunnel horizontally into the island. The direction of the electrode magnetization, via spin-orbit interaction, will thus affect the tunnelling density states in a way that can be sensed by the electrons tunnelling from the Au island. A microscopic analysis of the electronic structure of the quantum states involved in the tunneling process, based for example on first-principle calculations, is in our case not very feasible, since our thermally-grown Co electrodes are polycrystalline in character. In this case the tunneling density of states relevant for transport and its dependence of magnetization orientation is not simply related to the magneto-crystalline anisotropy[21] as for samples made of single crystals[10].

The TAMR found in our devices is very different from conventional TMR which is related to the relative magnetization orientation of two ferromagnetic electrodes separated by a thin tunnel barrier. In the latter case, the tunneling conductance is proportional to the product of the effective (tunneling) majority- and minority spin DOS at the Fermi level for the ferromagnetic electrodes[22]. In our case, an ordinary TMR would require the occurrence of spin accumulation on the central normal metal island, which is excluded by the fast spin-relaxation in Au samples of sizes of the order of 100 nm.

The TAMR discussed in the present paper is related to, but different from the TAMR reported in Ref 5 in that the anisotropy in the magnetoresistance reported in this previous work stems from DOS anisotropies in the valence band of GaAs induced by a broken cubic symmetry resulting from substitutional Mn doping.

In summary, we report on tunneling anisotropic magnetoresistance (TAMR) effects in nano-scaled Co-AlO$_x$-Au tunnel junctions. The anisotropy originates from a two-step magnetization reversal and an anisotropic tunneling density of states driven by spin-orbit interaction in the Co electrode. Our measurements open a new route to realize spin-valve-like effects in spintronics devices with only one ferromagnetic contact. Furthermore, our work shows the importance of careful investigations of the MR characteristics of all junctions in a spintronic device to draw correct conclusions from spin-transport measurements. We exemplify this point by analyzing the observed "TMR" in a ferromagnetic single-electron transistor and show that the anisotropic MR signal is in fact a TAMR signal in one of the junctions.

The authors thank Alexander N. Korotkov and David Haviland for fruitful discussions. The authors furthermore acknowledge financial support from Halmstad University, the Faculty of Natural Sciences at Kalmar University, the Swedish Research Council under Grant No: 621-2004-4439, the Swedish National Board for Industrial and Technological Development, the Office of Naval Research, the Knut and Alice Wallenberg Foundation, and the Swedish Foundation for Strategic Research.

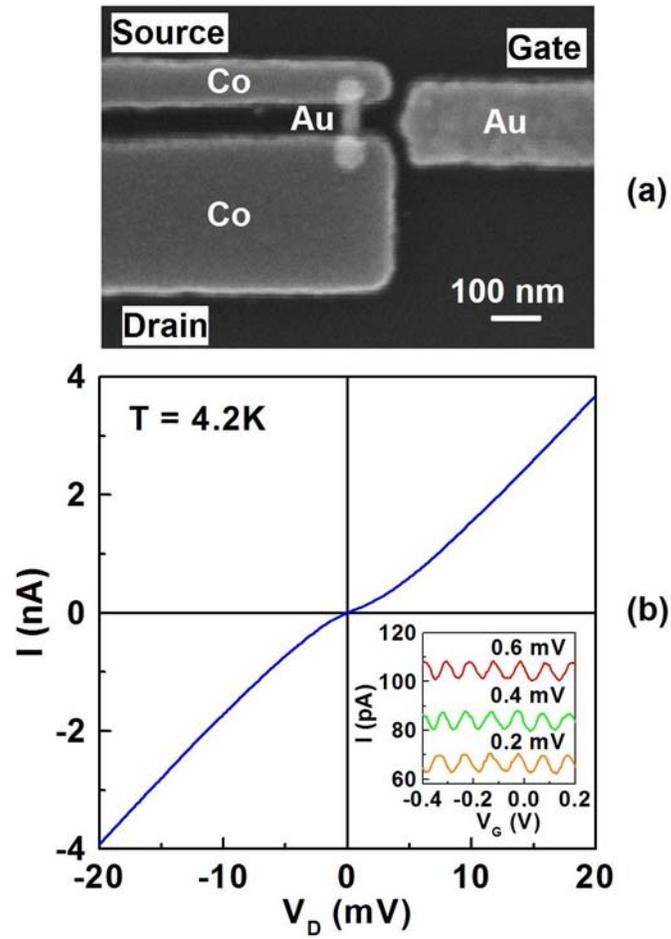

Fig.1. (a) SEM image of a Co/AlO$_x$/Au/AlO$_x$/Co device. (b) Nonlinear current-voltage (*I-V*) characteristics measured at 4.2K with no magnetic field. Lower right inset shows the gate-dependent current at different applied drain-source biases.

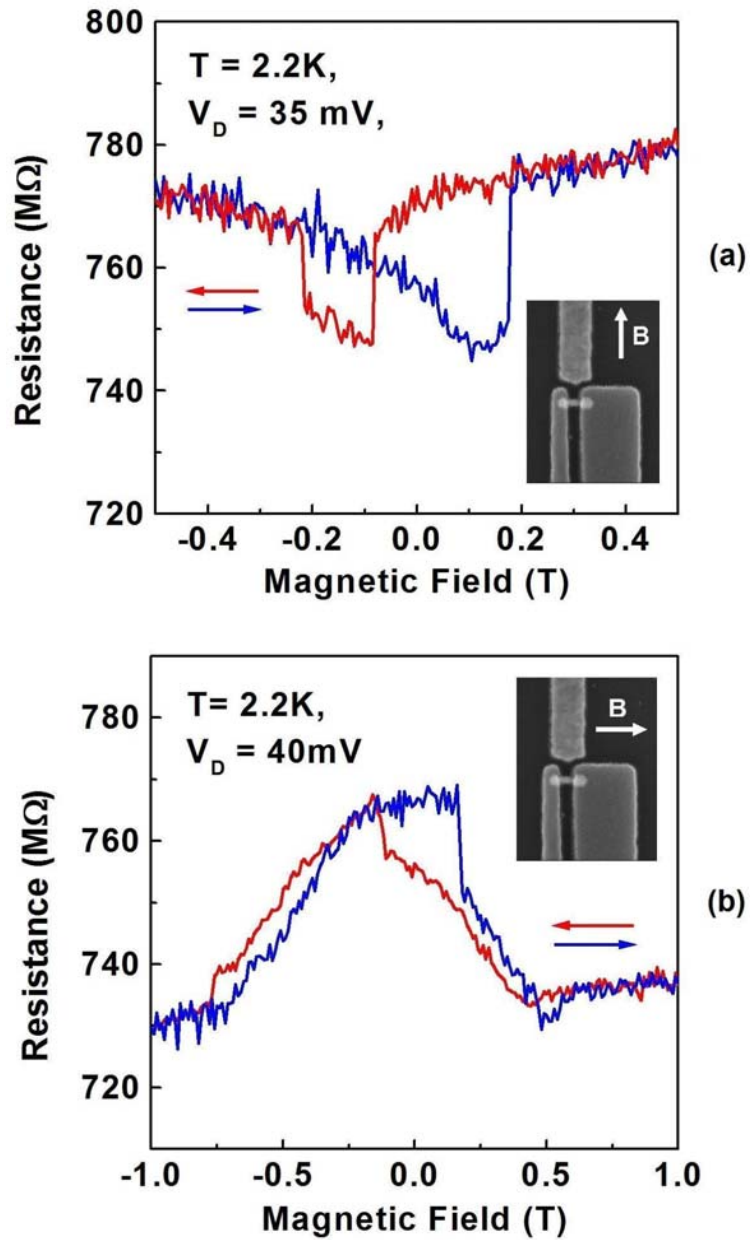

Fig. 2 Tunneling resistance versus magnetic field measured on Co/AlO$_x$/Au/AlO$_x$/Co ferromagnetic SETs at 2.2K with the field direction parallel (a) and perpendicular (b) to the long (easy) axis of the electrode, respectively. Insets show the SEM images and the field orientations.

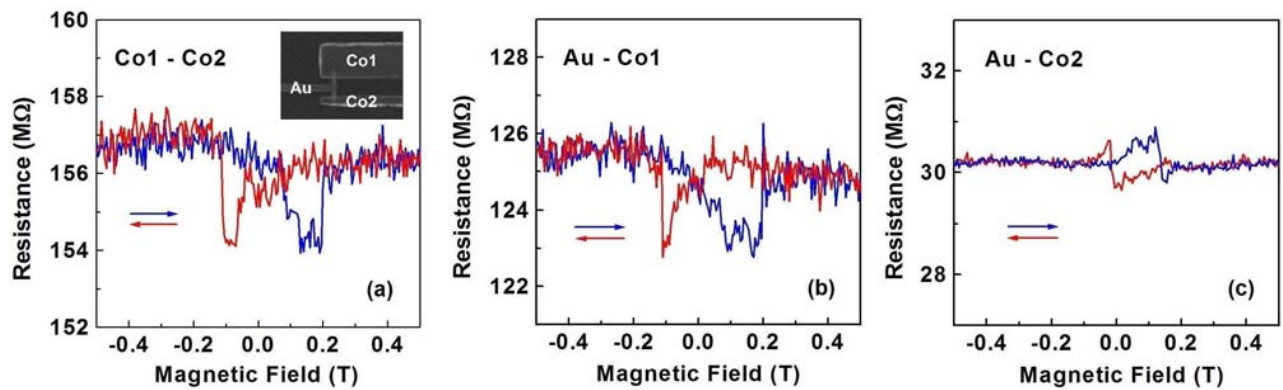

Fig. 3 Tunneling magnetoresistance versus magnetic field measured on a Co/AlO$_x$/Au/AlO$_x$/Co F-SET at 2.2K with the field applied parallel to the long (easy) axis of the electrode. The experimental results were obtained for (a) a Co1/AlO$_x$/Au/AlO$_x$/Co2 double junction, (b) a Co1/AlO$_x$/Au single junction and (c) a Co2/AlO$_x$/Au single junction, respectively. The inset in (a) shows a SEM image of the device.

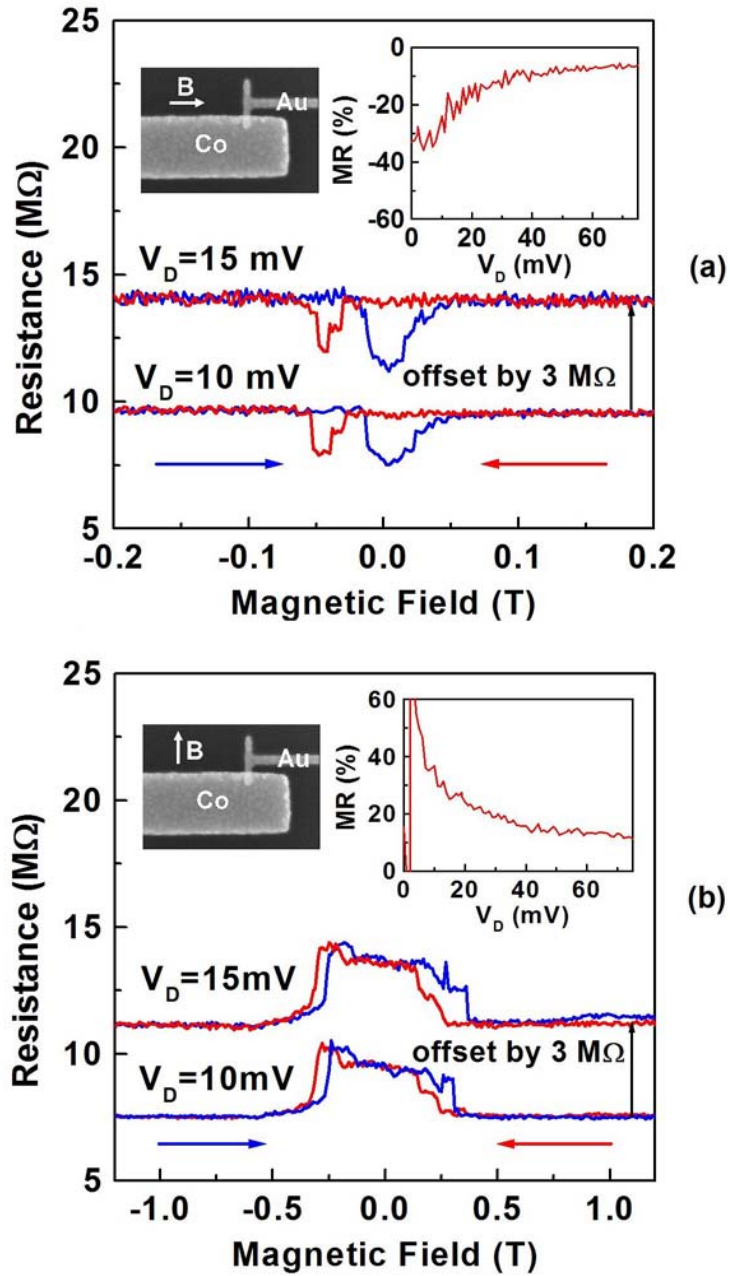

Fig. 4 Tunneling magnetoresistance versus magnetic field measured on a Co/AlO$_x$/Au junction at 4.2 K with the field direction applied parallel (a) and perpendicular (b) to the long (easy) axis of the Co electrode. Upper left insets show SEM images with indicated field directions. Upper right insets show the bias dependence of the tunneling magnetoresistance.